\documentclass[preprint,aps]{revtex4}
\usepackage{hyperref}
\usepackage{color}
\usepackage{graphicx}
\usepackage{bm}
\usepackage{amsmath}
\usepackage{hyperref}
\usepackage{enumerate}
\usepackage{upgreek}
\usepackage[subnum]{cases}

\newcommand{\ti}{ \tilde }

\newcommand{\pa}{ \partial }
\newcommand{\hb}{ \hbar }
\newcommand{\si}{ \sigma }

\newcommand{\ga}{ \gamma }
\newcommand{\la}{ \langle }
\newcommand{\ra}{ \rangle }

\newcommand{\al}{ \alpha }
\newcommand{\re}{ \text{Re} }

\newcommand{\erfc}{ \text{erfc} }
\newcommand{\pro}{ \text{Pr} }

\newcommand{\pr}{ \text{Pr} }

\begin{document} 
\title{Quantum backflow for dissipative two-identical-particle systems}
\author{S. V. Mousavi}
\email{vmousavi@qom.ac.ir}
\affiliation{Department of Physics, University of Qom, Ghadir Blvd., Qom 371614-6611, Iran}

\author{S. Miret-Art\'es}
\email{s.miret@iff.csic.es}
\affiliation{Instituto de F\'isica Fundamental, Consejo Superior de Investigaciones Cient\'ificas, Serrano 123, 28006 Madrid, Spain}

\begin{abstract}
In this work, dissipative quantum backflow is studied for a superposition of two stretched Gaussian wave packets and two 
identical spinless particles within the Caldirola-Kanai framework.  Backflow is mainly an interference process and 
dissipation is not able to suppress it in the first case. For two identical spinless particles, apart from interference terms, the 
symmetry of the wave function seems to be crucial in this dynamics. The combined properties make  bosons display this effect, even for the 
dissipative regime but, for fermions, backflow is not exhibited in any regime, dissipative and nodissipative. The anti-symmetric character 
of the corresponding wave function seems to be strong enough to prevent it. 
For bosons, backflow is also analyzed in terms of fidelity of one-particle states which is a well-known property of two 
quantum states. At very small values of fidelity, this effect is not seen even for bosons. 

\end{abstract}

\maketitle



\section{Introduction}

One of the purely quantum effects far less known that the tunneling effect is the so-called quantum backflow which is
a classically forbidden phenomenon where a positive-momentum wave packet along time may display negative fluxes during 
some time intervals and at some positions. This counter-intuitive effect takes place when an initial ensemble of free particles described by 
one-dimensional wave function, partially located in the negative axis of the coordinate and possessing only positive momenta, displays a 
non-decreasing probability of remaining in the negative region during certain periods of time. It was first described by Allcock \cite{allcock} 
when studying arrival times in quantum mechanics. Bracken and Melloy \cite{BrMe-JPA-1994} carried out subsequently  a systematic and 
detailed study  by providing an upper limit to the probability which can flow back from positive to negatives values of the coordinate 
to be around 0.04. 

Implications of backflow on quantum concepts of ``perfect absorption" and ``arrival time detection" has been addressed 
in \cite{MPaLe-PLA-1999}. Backflow is related to the interference of plane waves. It has been shown that a superposition of two Gaussian 
wave packets displays backflow but  not a single Gaussian wave packet \cite{Yearsley-2012}. In other words, for free motion there is no 
{\it quantum} backflow for a single Gaussian wave packet. 

Furthermore, as far as we know, no experimental evidence of this effect has been reported yet, though a feasible 
experimental scheme based on imprinting the backflow on a Bose-Einstein condensate and detecting it by a usual density measurement 
has been proposed \cite{PaToMuMO-PRA-2013}.
The maximum amount of probability for backflow occurring in general over any finite time interval is independent on the 
time interval, particle mass and Planck constant, talking about a new dimensionless quantum number. The same 
authors \cite{BrMe-AP-1998} also studied this effect in the presence of a constant field and in relativistic quantum mechanics 
from the Dirac equation. Interestingly enough, they  formulated the probability flow in terms of an eigenvalue problem of the flux operator. 
Following these studies, optimization numerical problems were reported by Penz et  al. \cite{Penz-2006}.
Superoscillations \cite{Berry-2006} and weak values \cite{Berry-2010}  in this context were considered by Berry. 
Yearsley et al. \cite{Yearsley-2012, Yearsley-2013} analyzed and discussed the classical limit as well as
some specific measurement models. Following the work by Bracken and Melloy, Albarelli {\em et al.} considered
the notion of nonclassicality arising from the backflow effect and analyzed its relationship with the negativity of the Wigner 
function \cite{Albarelli-2016}. 
Backflow has also been studied under the presence of a constant field and shown that it is mathematically equivalent to the 
problem of diffraction in time, \cite{Mo-PR-1952} for particles initially confined to a semi-infinite line and expanding them 
later on in free space \cite{Gu-PRA-2019}.
It has also been  extended to scattering problems in short-range potentials where it has been shown that this effect is a universal 
quantum one \cite{BoCaLe-PRA-2017}. Very recently, this effect has been studied for many-particles systems by Barbier \cite{Ba-arXiv-2020}.

Curiously, very few studies have been carried out in the context of open quantum systems. 
Yearsley dealt with the arrival time problem in the framework of decoherent histories for a particle coupled to an environment \cite{Ye-PRA-2010}. 
Recently, we have analyzed the dynamics of backflow in terms of dissipation and  addressed this issue within the Caldirola-Kanai
(CK) framework \cite{MoMi-EPJP-2020-2}.
Backflow has been shown to be reduced with dissipation. 
Its classical limit within the context of 
the classical Schr\"odinger equation \cite{Richardson-2014} has been also reported. 

Although the effect reported in \cite{MoMi-EPJP-2020-2} within the context of the Caldeira-Leggett framework is backflow, it is not 
{\it quantum} backflow as the one analyzed here. This is due to the fact that the interaction of the system with the environment makes that
the momentum distribution of the system displays positive and negative values during its time evolution \cite{erratum-2020}.

%
%
Following this previous work \cite{MoMi-EPJP-2020-2}, we have tackled the same theoretical analysis within the CK framework
by extending  the study to more general stretching Gaussian wave packets and to systems of two identical spinless particles, where the 
symmetry of the wave function adds a new ingredient to this effect. For the initial parameters chosen in this work,
bosons display backflow even for the dissipative case where decoherence leads to behave bosons, with symmetric wave function, as 
distinguishable particles. However,  fermions do not exhibit backflow even in the non-dissipative regime, the anti-symmetric property of their 
wave function being strong enough to prevent it in spite of  displaying interference. The general validity of these findings should be questioned
as far as a systematic study of the parameter space is carried out.
For bosons, backflow has also been analyzed in terms of fidelity which it is a measure of similarity of two pure one-particle states forming
the symmetric wave function. At very small values of fidelity, this effect is not seen even for bosons.

This work is organized as follows. In Section II, dissipative one-particle quantum backflow in the CK approach is analyzed.
Section III is devoted to extend this effect to two identical spinless particles, bosons and fermions, under the presence of dissipation.
In Section V, some numerical results are presented and discussed. Finally, a summary and some conclusions are drawn in Section V
at the end of this work.

\section{Dissipative one-particle quantum backflow in the Caldirola-Kanai approach}

In this section we briefly review dissipative quantum backflow for one-particle systems in the CK framework \cite{Caldirola-Kanai}. 
In this context, and for one-dimensional problems, the system is described by 
%
\begin{eqnarray} \label{eq: CK}
i \hb \frac{\pa }{\pa t} \psi(x, t)  &=& \bigg[ - e^{-2\ga t} \frac{\hb^2}{2m}  \frac{\pa^2}{\pa x^2}  + e^{2\ga t} V(x) \bigg] \psi(x, t)  
\end{eqnarray}
$V(x)$ being the external potential and $\ga$ the damping constant.  Along this work, we are going to consider only free motion, that is, $V(x)=0$. The canonical momentum $ p = - i \hb \pa / \pa x $ in this equation
fulfils the standard commutation relation $ [x, p] = i\hb $ and the physical momentum $ P $  is defined through the relation 
\begin{eqnarray} \label{eq: phy_mom}
P &=& e^{-2\ga t} p .
\end{eqnarray}

The probability current density $j(x, t)$ fulfilling the continuity equation
\begin{eqnarray} \label{eq: con_CK}
\frac{\pa |\psi(x, t)|^2}{\pa t} + \frac{\pa j(x, t)}{\pa x}  &=& 0 
\end{eqnarray}
is given by
\begin{eqnarray} \label{eq: pcd_CK}
j(x, t) &=& \frac{\hb}{m} \text{Im} \left\{ \psi^* \frac{\pa \psi}{\pa x}  \right\} e^{-2\ga t}   .
\end{eqnarray}
The probability that the particle can  be found in the region $(- \infty, 0]$ is
\begin{eqnarray} \label{eq: prob1}
\pro(x \leq 0,t) &=& \int_{-\infty}^{0}  |\psi(x, t)|^2  \, dx
\end{eqnarray}
and according to the continuity equation (\ref{eq: con_CK})
\begin{eqnarray} \label{eq: prob2}
\frac{d}{dt} \pro(x \leq 0,t)  &=& - j(0,t) 
\end{eqnarray}
provided that $j(- \infty,0) = 0$. If $j(0,t) \geq 0$ for $t \geq 0$ then $\pro(x \leq 0,t)$ is a decreasing function with time. 
However, if there are some time intervals where $j(0,t)$ is negative, the corresponding  probability  increases 
and we have the hallmark of backflow, whenever only  positive physical momenta contribute to the time-dependent wave function.

The wave function in the canonical-momentum space is given by the Fourier transform of the configuration space wave function,
\begin{eqnarray}
\ti{\psi}(p, t) &=& \frac{1}{ \sqrt{2\pi \hb} } \int dx ~ e^{-i p x / \hb} \psi(x, t)    ,
\end{eqnarray}
and the corresponding equation is given by
\begin{eqnarray} \label{eq: CK_fourier}
i \hb \frac{\pa }{\pa t} \ti{\psi}(p, t)  &=&  e^{-2\ga t} \frac{p^2}{2m} \ti{\psi}(p, t)  + e^{2\ga t} 
\mathcal{F}[V(x) {\psi}(x, t)] 
\end{eqnarray}
where $ \mathcal{F} $ means the Fourier transform of its argument. The  canonical momentum distribution function is expressed as
\begin{eqnarray} \label{eq: ca_mom_dis}
\ti{\rho}(p, t)  &=&  | \ti{\psi}(p, t) |^2     
\end{eqnarray}
and due to the one-to-one relationship (\ref{eq: phy_mom}) between the physical and canonical momentum, the corresponding probability densities  
are related by
\begin{eqnarray} 
\ti{\uprho}(P, t) dP &=& \ti{\rho}(p, t) dp     .
\end{eqnarray}
Thus, we have that
\begin{eqnarray} \label{eq: phys_mom_dis_1}
\ti{\uprho}(P, t) &=& \ti{\rho}(p, t) \frac{dp}{dP} \bigg|_{p= P e^{2\ga t}} = e^{2\ga t} \ti{\rho}(e^{2\ga t} P, t) 
\end{eqnarray}
and the probability of finding a negative value for the physical momentum is 
\begin{eqnarray} \label{eq: prob_pneg}
\pr(P<0, t) &=& \int_{-\infty}^0 dP ~ \ti{\uprho}(P, t) 
= \int_{-\infty}^0 dP ~ e^{2\ga t} \ti{\rho}(e^{2\ga t} P, t)
= \int_{-\infty}^0 dp ~ \ti{\rho}(p, t) 
\nonumber \\
&=& \pr(p<0, t)
\end{eqnarray}
which is just the probability of obtaining a negative value for the canonical momentum. In backflow studies,
one should make sure that the  wave packet used to describe the particles remains with positive momenta along time.


The solution of Eq. (\ref{eq: CK_fourier}) for $V(x)=0$ is
\begin{eqnarray} \label{eq: F_free}
\ti{\psi}(p, t)  &=&  \ti{\psi}_0(p) ~ \exp \left[ - \frac{i}{\hb} \frac{p^2}{2m} \uptau(t) \right]
\end{eqnarray}
where $ \ti{\psi}_0(p) = \ti{\psi}(p, 0) $ and
\begin{eqnarray} \label{eq: tau}
\uptau(t) &=& \frac{1-e^{-2\ga t}}{2\ga}   .
\end{eqnarray}
From Eq. (\ref{eq: ca_mom_dis}) it is then seen that the canonical momentum distribution function is independent on time. With respect to 
Eq. (\ref{eq: prob_pneg}), this means that if initially the contribution of negative momenta is negligible, it remains so along time. 
The friction force acts against the free motion but this force ultimately stops the particle and  does not reverse the  motion. 

For the non-minimum-uncertainty-product or stretched Gaussian wave packet defined as
\begin{eqnarray} \label{eq: psit_0}
\ti{\psi}_0(p)  &=&  \frac{1}{(2\pi \si_p^2)^{1/4}} \exp \left[ - (1 + i \eta) \frac{ (p-p_0)^2 }{ 4\si_p^2 } - \frac{i}{\hb} x_0~p \right]
\end{eqnarray}
and from Eq. (\ref{eq: prob_pneg}), one obtains the corresponding probability in terms of the complementary error function
\begin{eqnarray} \label{eq: prob_pneg_free}
\pr(P<0, t) &=& \frac{1}{2} \erfc \left[ \frac{p_0}{\sqrt{2} \si_p} \right]
\end{eqnarray}
which as mentioned above is time-independent. The initial values $\si_p$ and $p_0$ should be chosen in 
such a  way that this probability is nearly zero. The Fourier transform of Eq. (\ref{eq: psit_0}) yields
\begin{eqnarray} \label{eq: psi_0}
\psi_0(x)  &=&  \frac{1}{(2\pi \si_0^2(1+\eta^2))^{1/4}} \exp \left[ - \frac{(x-x_0)^2}{ 4\si_0^2( 1+\eta^2 ) } + \frac{i}{\hb} p_0(x-x_0) \right]
\end{eqnarray}
where 
\begin{eqnarray}
\si_0 &=& \frac{\hb}{2\si_p} 
\end{eqnarray}
and $ \eta $ is called the stretching parameter since $\Delta x = \sqrt{ \la x^2 \ra - \la x \ra^2 } = \si_0 \sqrt{1+\eta^2}$.


For a linear potential such as
\begin{eqnarray} \label{eq: lin_pot}
V(x) &=&  - m g ~ x
\end{eqnarray}
where $ g $ plays the role of an acceleration, Eq. (\ref{eq: CK_fourier}) is now expressed  as
\begin{eqnarray} \label{eq: CK_force}
i \hb \frac{\pa }{\pa t} \ti{\psi}(p, t)  &=&  \left[ e^{-2\ga t} \frac{p^2}{2m} - e^{2\ga t} i \hb m g \frac{\pa}{\pa p} \right] \ti{\psi}(p, t)    .
\end{eqnarray}
The solution of this equation with the initial condition (\ref{eq: psit_0}) leads to
\begin{eqnarray} \label{eq: psit_t}
\ti{\psi}(p, t)  &=&  \frac{1}{(2\pi \si_p^2)^{1/4}} \exp \left[ - \frac{s_t}{2\hb\si_p} (p-p_t e^{2\ga t})^2 - \frac{i}{\hb} p x_t + \frac{i}{\hb} \mathcal{A}_t \right]
\end{eqnarray}
where 
\begin{numcases}~
s_t = \frac{\hb}{2\si_p} \left[ 1 + i \left( \frac{2\si_p^2}{m\hb} \uptau(t) + \eta\right) \right]   \label{eq: st} \\
x_t = x_0 + \frac{p_0}{m} \uptau(t) + g \frac{ 2\ga t - 1 + e^{-2\ga t} }{ 4 \ga^2 }  \label{eq: xt} \\
p_t = m \dot{x}_t = p_0 e^{-2\ga t} + m g \uptau(t) \label{eq: pt} \\
\mathcal{A}_t = \frac{p_0^2}{2m} \uptau(t) + g \left( p_0 \frac{-1+\cosh(2\ga t)}{2\ga^2} + m x_0 e^{2\ga t} \uptau(t) \right) + 
m g^2 \frac{ 4 + (4\ga t-3) e^{2\ga t} - e^{-2\ga t} }{ 16\ga^3  }  \nonumber \\  \label{eq: At} 
\end{numcases}
$s_t$, $x_t$, $p_t$ and $\mathcal{A}_t$ being the complex width of the wave packet in configuration space, its center and kick 
momentum and the classical action, respectively.
By replacing Eq. (\ref{eq: psit_t}) into Eq. (\ref{eq: ca_mom_dis}) and using Eq. (\ref{eq: phys_mom_dis_1}), one obtains
\begin{eqnarray} \label{eq: phys_mom_dis_linpot}
\ti{\uprho}(P, t) &=& \frac{1}{\sqrt{2\pi} \si_p(t)} \exp \left[ - \frac{ (P-p_t)^2 }{ 2\si^2_p(t) } \right] 
\end{eqnarray}
for the physical momentum distribution function where we have defined
\begin{eqnarray} \label{eq: sigmapt}
\si_p(t) &=& e^{-2\ga t} \si_p      .
\end{eqnarray}
Distribution (\ref{eq: phys_mom_dis_linpot}) becomes ultimately a Dirac delta function centered at $ p_{\infty} = m g / 2\ga $. 
This is physically an acceptable result since the friction force acts against the constant force $ m g $; classically, particles ultimately take the same velocity 
$ v_{\infty} = g / 2\ga $ and thus the width of the momentum distribution is zero in the limit $ t \rightarrow \infty $.
Then, the probability of obtaining a negative value in a measurement of the physical momentum is
\begin{eqnarray} \label{eq: prob_pneg_linpot}
\pr(P<0, t) &=& \frac{1}{2} \erfc \left[ \frac{p_t}{\sqrt{2} \si_p(t)} \right]
= \frac{1}{2} \erfc \left[ \frac{p_0}{\sqrt{2} \si_p} + \frac{m g}{\sqrt{2} \si_p} \frac{ e^{2\ga t}-1 }{ 2\ga } \right]
\end{eqnarray}
where in the second equality we have used Eqs. (\ref{eq: pt}) and (\ref{eq: sigmapt}). The argument of the complementary error function 
increases with time i.e., $ \pr(P<0, t) $ is a decreasing function of time. 


As is well known, quantum backflow does not occur for a single Gaussian wave packet; one needs a superposition of at least two Gaussians. 
We take the initial momentum space wave function as a superposition of two stretched Gaussians with the same width but different kick 
momenta 
\begin{eqnarray} \label{eq: wf0_momentum}
\ti{\psi}_0(p) &=& N \frac{1}{(2\pi \si_p^2)^{1/4}} \left\{
\exp \left[ - (1 + i \eta) \frac{ (p-p_{0a})^2 }{4\si_p^2} \right]
+ \al e^{i\theta} \exp \left[ - (1 + i \eta) \frac{ (p-p_{0b})^2 }{4\si_p^2} \right]
\right \}, \nonumber \\
\end{eqnarray}
where $ N $, the normalization constant,  $\al$ and $\theta$ are all real numbers with
\begin{eqnarray} \label{eq: N_cons}
N &=& \left( 1 + \al^2 + 2 \al \cos \theta ~ \exp\left[ - \frac{ (p_{0a}-p_{0b})^2 }{ 8 \si_p^2} 
( 1 + \eta^2 ) \right] \right)^{-1/2}    .
\end{eqnarray}
The negative momentum probability, being independent on time, is obtained from Eq. (\ref{eq: prob_pneg})
\begin{eqnarray} \label{eq: prob_p<0}
\pr(P<0, t) &=& \int_{-\infty}^0 dp | \ti{\psi}_0(p) |^2 
\nonumber \\
&=& 
\frac{1}{2} N^2 \bigg\{ 
\erfc \left[  \frac{p_{0a} }{ \sqrt{2} \si_p } \right] 
+ \alpha^2 \erfc \left[ \frac{p_{0b} }{ \sqrt{2} \si_p } \right]+ \al~ e^{- (p_{0a}-p_{0b})^2(1 + \eta^2) / 8\si_p^2}
\nonumber \\
& & 
\qquad \times \left( e^{i\theta} \erfc \left[ \frac{ p_{0a}+p_{0b}-i \eta(p_{0a}-p_{0b}) }{ 2\sqrt{2}\si_p } \right] + \textbf{c.c.} \right) \bigg\}   ,
\end{eqnarray}
where c.c. means the complex conjugation of the first term in  parenthesis. 
In the free propagation, the configuration space wave function is given by  
\begin{eqnarray} \label{eq: wf_sup_t}
\psi(x, t) &=& N ( \psi_a(x, t) + \al e^{i\theta} \psi_b(x, t) )   
\end{eqnarray}
$\psi_a$ and $\psi_b$ being two stretched Gaussian wave packets expressed as ($i=a,b$)
\begin{eqnarray} \label{eq: Gausst}
\psi_i(x, t) &=& \frac{1}{(2\pi s_t^2)^{1/4}} \exp \left[ - \frac{\si_p (x-x_{ti})^2}{2\hb s_t} + i \frac{p_{0i}}{\hb} (x-x_{ti}) + \frac{i}{\hb} \mathcal{A}_{ti} \right]
\end{eqnarray}
where $s_t$, $x_{ti}$ and $\mathcal{A}_{ti}$ are  given by Eqs. (\ref{eq: st}), (\ref{eq: xt}) and (\ref{eq: At}) respectively and 
where  the two conditions $g =0 $ and $p_0=p_{0i}$ have been imposed. 
By computing the probability density and then integrating over the negative  half $x$ axis, one has
\begin{eqnarray} \label{eq: prob_x<0_CK}
\mathcal{P}(t) &=& \frac{1}{2} N^2 \bigg\{ 
\erfc \left[ \frac{ x_{ta} }{\sqrt{2} \si_t}  \right]
+ \al^2 \erfc \left[ \frac{ x_{tb} }{\sqrt{2} \si_t} \right]
+ \al e^{-(p_{0a} - p_{0b})^2(1+\eta^2) / 8\si_p^2 } \nonumber \\
& & \qquad \times \bigg( e^{i\theta} \erfc \left[ \frac{d(t)}{\sqrt{2}\si_t} \right] 
+  \text{c.c.} \bigg)
\bigg \}
\end{eqnarray}
for the probability of remaining in the region $x<0$ with
\begin{numcases}~
d(t) = \frac{ p_{0a} + p_{0b} }{2m} \uptau(t) 
- i \left( \frac{m \hb}{2\si_p^2}(1+\eta^2) + \eta \uptau(t) \right) \frac{ p_{0a} - p_{0b} }{2m}
\\
\si_t = |s_t| = \frac{\hb}{2\si_p} \sqrt{ 1 + \left( \frac{2\si_p^2}{m\hb} \uptau(t) + \eta\right)^2 }   .
\end{numcases}
Note that the arguments of the first two complementary error functions are increasing functions of time. Thus, the last two terms, i.e. 
the interference terms, are responsible for dissipative quantum backflow.

\section{Dissipative quantum backflow for two identical particles}

In the CK framework, a dissipative two-particle system is described by the two-particle equation
\begin{eqnarray} \label{eq: CK_2p}
i \hb \frac{\pa }{\pa t} \Psi(x_1, x_2, t)  &=& \bigg[ e^{-2\ga t} \left( - \frac{\hb^2}{2m_1}  \frac{\pa^2}{\pa x_1^2} - \frac{\hb^2}{2m_2}  \frac{\pa^2}{\pa x_2^2} \right)  + e^{2\ga t} U(x_1, x_2) \bigg] \Psi(x_1, x_2, t)  
\end{eqnarray}
with $ \Psi(x_1, x_2, t) $ being the two-particle configuration-space wave function and $U(x_1,x_2)$, the interparticle interaction. The
canonical-momentum space wave function, being the Fourier transform of the configuration space wave function, is given by
\begin{eqnarray}
\ti{\Psi}(p_1, p_2, t) &=& \frac{1}{ 2\pi \hb} \int dx_1 \int dx_2~ e^{-i p_1 x_1 / \hb} e^{-i p_2 x_2 / \hb} \Psi(x_1, x_2, t)
\end{eqnarray}
which fulfils as before the wave equation
\begin{eqnarray} \label{eq: 2CK_fourier}
i \hb \frac{\pa }{\pa t} \ti{\Psi}(p_1, p_2, t)  &=&  e^{-2\ga t} \left( \frac{p_1^2}{2m_1} + \frac{p_2^2}{2m_2} \right) \ti{\Psi}(p_1, p_2, t) + e^{2\ga t} 
\mathcal{F}[U(x_1, x_2) \Psi(x_1, x_2, t)]    .
\end{eqnarray}
%
In general, for a system of two identical particles, the wave function describing the system must have a given symmetry. 
If $  \chi_0  $ and $  \phi_0  $ are the initial ($t=0$) one-particle states with $\ti{\chi}_0(p)$ and $\ti{\phi}_0(p)$ the corresponding 
momentum space wave functions, then the initial wave function in this space is expressed as  
\begin{eqnarray} \label{eq: 2p-psi0_F}
\ti{\Psi}_{\pm}(p_1, p_2, 0) &=& \mathcal{N}_{\pm} ( \ti{\chi}_0(p_1) \ti{\phi}_0(p_2) \pm \ti{\phi}_0(p_1) \ti{\chi}_0(p_2) )
\end{eqnarray}
where $+$ and $-$ refer to bosons and fermions, respectively. 
When particles do not interact with each other, but do interact with  an external  potential i.e., $ U(x_1, x_2) = V(x_1) + V(x_2) $, 
then from the linearity of the CK wave equation, the time-dependent symmetric and anti-symmetric 
solutions can be written as
\begin{eqnarray} \label{eq: 2p-psit_F}
\ti{\Psi}_{\pm}(p_1, p_2, t) &=& \mathcal{N}_{\pm} ( \ti{\chi}(p_1, t) \ti{\phi}(p_2, t) \pm \ti{\phi}(p_1, t) \ti{\chi}(p_2, t) )
\end{eqnarray}
where $\ti{\chi}(p, t)$ and $\ti{\phi}(p, t)$ fulfill the corresponding one-particle CK wave equation. Apart from a phase factor, 
the normalization constants $ \mathcal{N}_{\pm} $ are given by
\begin{eqnarray} \label{eq: normalization}
\mathcal{N}_{\pm} &=& \frac{1}{ \sqrt{ 2(1 \pm | \la \chi | \phi \ra |^2)  } }
\end{eqnarray}
where it is assumed that the one-particle wave functions $ \chi_0 $ and $  \phi_0 $ are normalized. We have also used the fact that the 
overlaping  $ \la \chi(t) | \phi(t) \ra $ is independent on time.
 
From Eq. (\ref{eq: 2p-psit_F}), the two-particle {\it canonical} momentum distribution function is given by
\begin{eqnarray}  \label{eq: 2p-fourier}
| \ti{\Psi}_{\pm}(p_1, p_2, t) |^2 &=& 
\mathcal{N}_{\pm}^2 \bigg( |\ti{\chi}(p_1, t)|^2 |\ti{\phi}(p_2, t)|^2 + |\ti{\phi}(p_1, t)|^2 |\ti{\chi}(p_2, t)|^2 \nonumber \\
&\pm& \re\{ \ti{\chi}^*(p_1, t) \ti{\phi}(p_1, t) \ti{\phi}^*(p_2, t) \ti{\chi}(p_2, t) \}
 \bigg) 
\end{eqnarray}
and following the same analysis which yields  Eq. (\ref{eq: phys_mom_dis_1}), one has
\begin{eqnarray} \label{eq: phys_mom_dis_2}
\ti{\uprho}_{\pm}(P_1, P_2, t) &=& | \ti{\Psi}_{\pm}(p_1, p_2, t) |^2 \frac{dp_1}{dP_1} \bigg|_{p_1= P_1 e^{2\ga t}} \frac{dp_2}{dP_2} \bigg|_{p_2= P_2 e^{2\ga t}} \nonumber \\
&=& e^{4 \ga t} | \ti{\Psi}_{\pm}(P_1 e^{2\ga t}, P_2 e^{2\ga t}, t) |^2
\end{eqnarray}
for the physical momentum distribution function. For dissipative quantum backflow, we make to be sure again 
that negative-momentum contributions to the wave function are negligible along time. 
To this end, we first compute the probability that both outcomes are {\it positive} in a simultaneous 
measurement of physical momenta 
\begin{eqnarray} \label{eq: prob_ppos_2}
\pr(P_1>0, P_2>0, t) &=& \int_0^{\infty} dP_1 \int_0^{\infty} dP_2 ~ \ti{\uprho}(P_1, P_2, t) 
= \pr(p_1>0, p_2>0, t) 
\end{eqnarray}
and from the relation $ \pr(P_1>0, P_2>0, t) + \pr(P_1>0, P_2<0, t) + \pr(P_1<0, P_2>0, t) + \pr(P_1<0, P_2<0, t) = 1 $ one has
\begin{eqnarray} \label{eq: prob_one_pneg}
\ti{\mathcal{P}}(t) &=& 1 - \pr(P_1>0, P_2>0, t)
\end{eqnarray}
which gives the probability of obtaining {\it at least} a negative value in a simultaneous momentum measurement.

For {\it freely} propagating wave packets, one sees from Eq.(\ref{eq: F_free}) that the distribution (\ref{eq: 2p-fourier}) is independent on time 
i.e., $ | \ti{\Psi}_{\pm}(p_1, p_2, t) |^2 = | \ti{\Psi}_{\pm}(p_1, p_2, 0) |^2 $. Thus, the probability $ \ti{\mathcal{P}}(t) $ remains unchanged.
The configuration space wave function is then obtained by taking the Fourier transform of Eq. (\ref{eq: 2p-psit_F}) which yields
\begin{eqnarray} \label{eq: 2p-psit}
\Psi_{\pm}(x_1, x_2, t) &=& \mathcal{N}_{\pm} ( \chi(x_1, t) \phi(x_2, t) \pm \phi(x_1, t) \chi(x_2, t) )
\end{eqnarray}
and the probability of finding at least one particle in the negative half-space is given by \cite{Ba-arXiv-2020}
\begin{eqnarray} \label{eq: prob_one_neg}
\mathcal{P}_{\pm}(t) &=& \mathcal{P}_{\pm, \text{nn}}(t) + 2 \mathcal{P}_{\pm, \text{pn}}(t)
\end{eqnarray}
where $ \mathcal{P}_{\pm, \text{nn}}(t) $ displays the probability of finding both particles in the negative half space 
and $ \mathcal{P}_{\pm, \text{pn}}(t) $ is the corresponding probability of finding one particle in the positive half space but 
the other in the negative part. Due to the symmetry of the wave function one observes that 
$ \mathcal{P}_{\pm, \text{pn}}(t) = \mathcal{P}_{\pm, \text{np}}(t) $ and from the normalization condition one has that 
\begin{eqnarray} \label{eq: prob_one_neg2}
\mathcal{P}_{\pm}(t) &=&  1 - \mathcal{P}_{\pm, \text{pp}}(t) = 1 - \int_0^{\infty} dx_1 \int_0^{\infty} dx_2 ~ | \Psi_{\pm}(x_1, x_2, t) |^2 .
\end{eqnarray}
The two one-particle states $\chi(x, t)$ and $\phi(x, t)$ appearing in 
Eq. (\ref{eq: 2p-psit}) are built from the superposed state Eq. (\ref{eq: wf_sup_t}) with the same $\alpha$ but different $\theta$; 
$\theta_{\chi}$ and $\theta_{\phi}$, respectively.

\section{Numerical Calculations}

Atomic units  ($\hb = m = 1$) are used along this work and calculations are only carried out in the absence of any interaction potential i.e., 
$ V = 0 $.
\begin{figure}
	\centering
	\includegraphics[width=12cm,angle=-0]{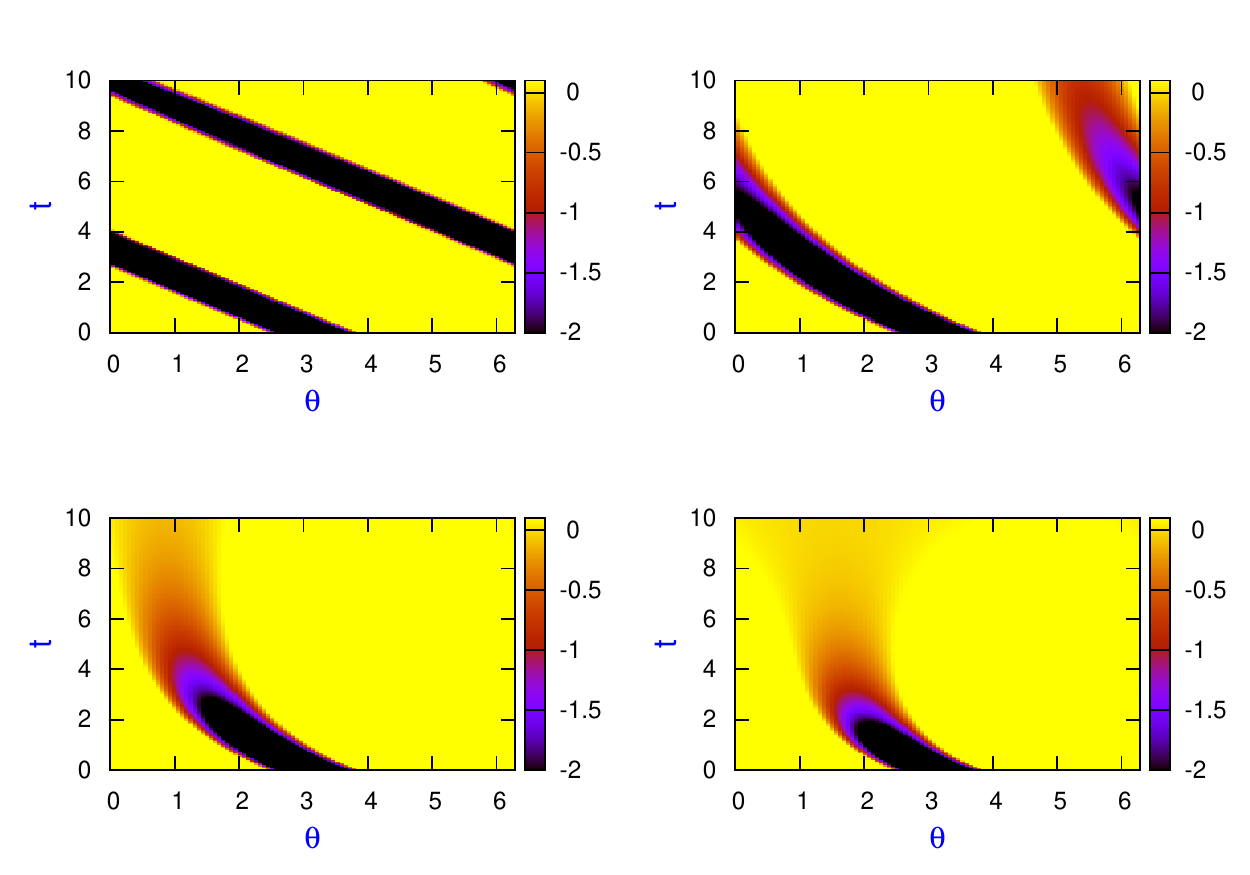}
	\caption{
		Density plots of the probability current density at the origin, magnified by a factor of 1000, for the one-particle superposed state Eq. (\ref{eq: wf_sup_t}), with the 
		component wave packets as minimum-uncertainty-product Gaussians and different damping constants:
		$ \ga=0 $ (left top panel), 
		$ \ga=0.1 $ (right top panel), $ \ga=0.2 $ (left bottom panel) and $ \ga=0.3 $ (right bottom panel). Parameters are chosen to be 
		$x_{0a} = x_{0b} = 0 $, $\si_{0a} = \si_{0b} = 10$, $ p_{0a}=1.4 $, $ p_{0b}=0.3 $ and $\al=1.9$. } 
	\label{fig: cur_bf} 
\end{figure}
Unless otherwise stated, the one-particle wave packets $ \psi_a $ and $ \psi_b $ are described by minimum-uncertainty-product Gaussian 
ones given initially by  Eq. (\ref{eq: psi_0}) 
with $ \eta=0 $, the same center $ x_0 = 0 $ and width $ \si_0 = 10 $, equivalently $\si_p=0.05$, but different kick momenta: 
$ p_{0a}=1.4 $ and $ p_{0b}=0.3 $. Density plots of the probability current density at the origin, $x=0$, magnified by a factor of 1000, 
have been depicted in Figure \ref{fig: cur_bf} for the one-particle superposed 
state Eq. (\ref{eq: wf_sup_t}) with $ \al = 1.9 $ and different damping constants: $ \ga=0 $ (left top panel), 
$ \ga=0.1 $ (right top panel), $ \ga=0.2 $ (left bottom panel) and $ \ga=0.3 $ (right bottom panel). 
With these parameters, the probability of obtaining a negative value in a momentum measurement given by Eq. (\ref{eq: prob_p<0}) is of the order of 
$ 10^{-10} $ for all values of $\theta$ and this probability is time-independent. 
According to the color gradient given in Figure \ref{fig: cur_bf}, at $t=0$ there is an interval around $ \theta = \pi $ where the probability 
current density is negative and the backflow effect should take place. This interval moves to smaller values of $\theta$ during the time 
evolution. 
In the time domain plotted i.e., $ t \in [0,10] $ and for a given value of $ \theta $ for which $ j(0,0) < 0 $, there are two backflow intervals 
for $\ga=0$ while it remains only one backflow interval with dissipation.

\begin{figure}
	\centering
	\includegraphics[width=12cm,angle=-0]{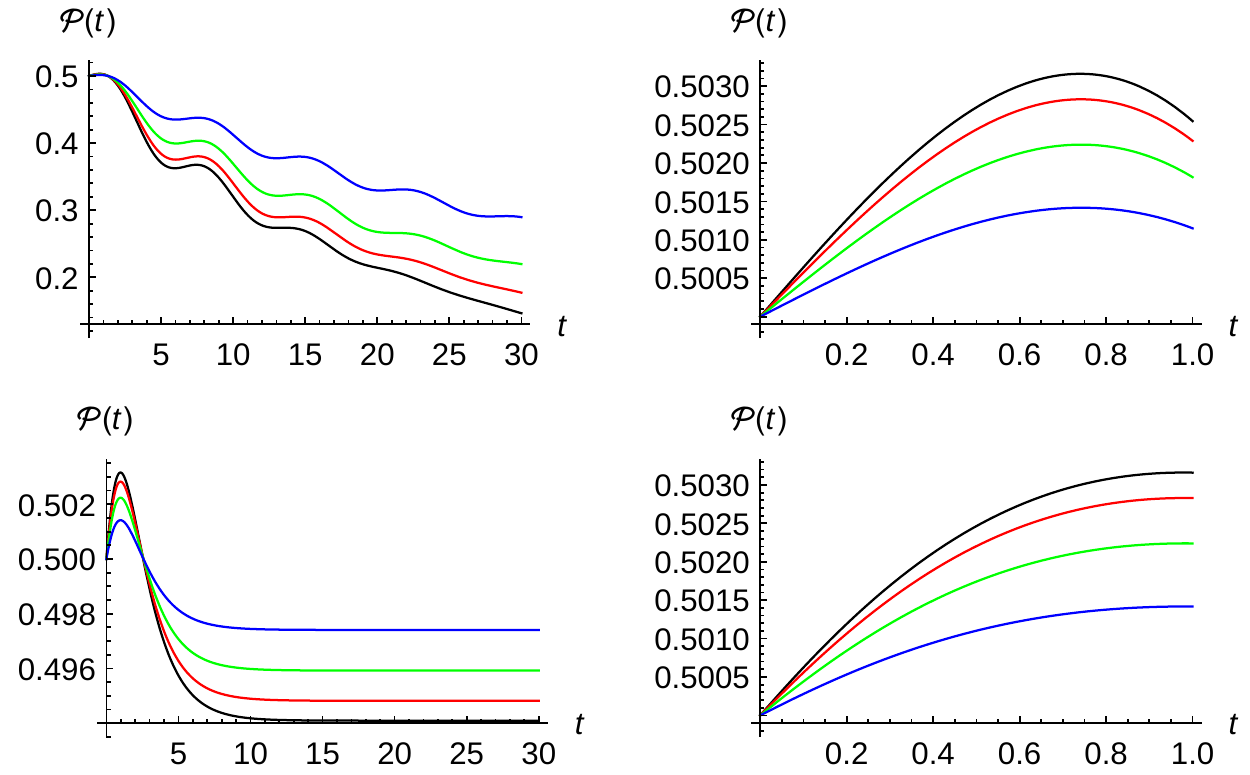}
	\caption{
		Probability  given by Eq. (\ref{eq: prob_x<0_CK}) of finding the particle described by the one-particle superposed state, 
		Eq. (\ref{eq: wf_sup_t}) composed of two stretched Gaussian wavepackets, in the negative half-space versus time for $\ga=0$ 
		(top plots) and $\ga=0.3$ (bottom plots) with different 
		values of the stretching parameter; $ \eta=0 $ (black curves), $ \eta=0.5 $ (red curves), $ \eta=1 $ (green curves) and $ \eta=2 $ (blue curves).
		Right panels are plots magnified around short times where backflow takes place.
		Parameters are chosen to be $x_{0a} = x_{0b} = 0 $, $\si_{0a} = \si_{0b} = 10$, $ p_{0a}=1.4 $, $ p_{0b}=0.3 $, $\al=1.9$ 
		and $\theta=\pi$.} 
	\label{fig: bf_eta} 
\end{figure}

In Figure \ref{fig: bf_eta}, the probability $ \mathcal{P}(t) $, Eq. (\ref{eq: prob_x<0_CK}),  of finding the particle described by the 
one-particle superposed state given by Eq. (\ref{eq: wf_sup_t}) composed of two stretched Gaussian wave packets, in the negative half-space, 
is plotted versus time for $\ga=0$ (top plots) and $\ga=0.3$ (bottom plots) and different values of the stretching parameter:
$ \eta=0 $ (black curves), $ \eta=0.5 $ (red curves), $ \eta=1 $ (green curves) and $ \eta=2 $ (blue curves). Here, $\alpha = 1.9 $ and 
$ \theta=\pi $; both wave packets $\psi_a$ and $\psi_b$ being stretched Gaussian wave packets with the same center 
$x_0=0$ and width $\si_0=10$ and different kick momenta $ p_{0a}=1.4 $ and $ p_{0b}=0.3 $. For these parameters, $ \pr(P<0, t) $, 
being time-independent, is of the order of $ 10^{-10} $ for all values of the stretching parameter considered.
As the left top panel shows, there are several backflow intervals for the non-dissipative dynamics where the first one starts around $t=0$. However, 
when dissipation is present (left bottom panel) only the first time interval backflow remains. Thus, backflow is not suppressed with dissipation; at least for our parameters. 
Furthermore, this probability goes to a stationary value due to the constant value of the wave packet width. The role of the stretching parameter 
is not negligible at all.
As seen in the right panels where a magnification at short times of the first backflow is seen,
the amount of backflow quantified by $ |\mathcal{P}(t_m) - \mathcal{P}(0)| $, $t_m$ being the time where 
$ \mathcal{P} $ is maximum, diminishes with the stretching parameter.
\begin{figure}
	\centering
	\includegraphics[width=12cm,angle=-0]{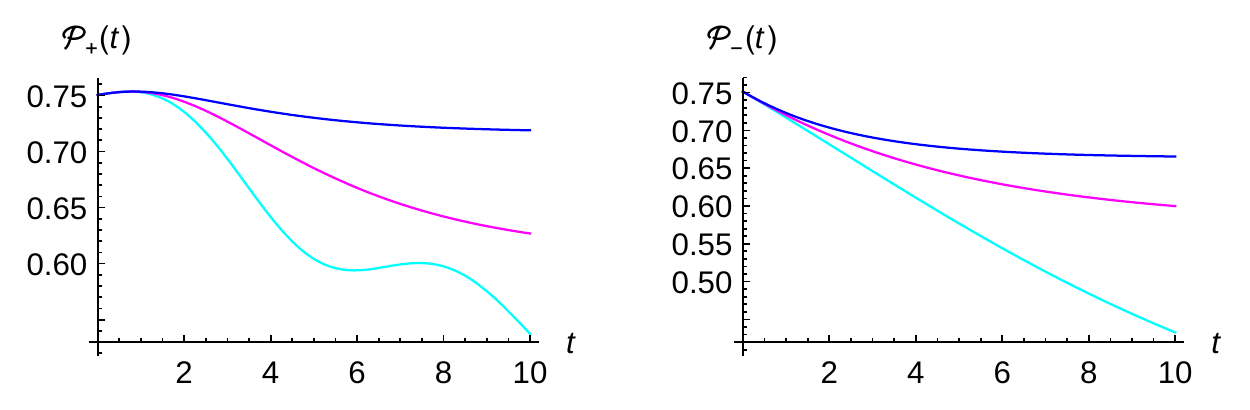}
	\caption{
		Probability of finding at least a particle in the negative half-space for bosons (left panel) and fermions (right panel) versus time for different values of damping constant; $ \ga=0 $ (cyan curves), $ \ga=0.1 $ (magenta curves) and $ \ga=0.2 $ (blue curves). Parameters are chosen to be $ \si_p=0.05 $,  $ p_{0b}=0.3 $, $ p_{0a}=1.4 $, $\al=1.9$, $\theta_{\chi}=\pi$ and $\theta_{\phi}=1.01 \pi$.} 
	\label{fig: bf_2p} 
\end{figure}
%

%
%

Now the next step is to analyze the backflow effect for two-identical-particle systems. For this goal, we have to choose the two one-particle states 
$\chi(x, t)$ and $\phi(x, t)$ appearing in Eq. (\ref{eq: 2p-psit}) in the form of the superposed state Eq. (\ref{eq: wf_sup_t}) with 
the same $\alpha$ but different $\theta$; $\theta_{\chi}$ and $\theta_{\phi}$, respectively. 
The remaining parameters defining the component Gaussian wave packets are the same as before. 
In Figure  \ref{fig: bf_2p},  the probability of finding at least a particle in the negative half-space is depicted for bosons 
($ \mathcal{P}_{+}(t) $,  left panel) and fermions  ($ \mathcal{P}_{-}(t) $, right panel) versus time for different values of the damping 
constant; $ \ga=0 $ (cyan curves), $ \ga=0.1 $ (magenta curves) and $ \ga=0.2 $ 
(blue curves). Parameters are chosen to be: $ \si_p=0.05 $,  $ p_{0b}=0.3 $, $ p_{0a}=1.4 $, $\al=1.9$, $\theta_{\chi}=\pi$ and 
$\theta_{\phi}=1.01 \pi$. According to this figure, with these parameters, backflow occurs only for bosons, not for fermions. 
When $\gamma = 0$ and only positive physical momenta are present, $ \mathcal{P}_{+,\text{pp}}(t) $ is an increasing oscillatory 
function with time due to interference terms and the symmetry of the wave function  favors that the two particles 
remain close each other in the positive as well as the negative part of the $x$-axis.
According to Eq. (\ref{eq: prob_one_neg2}), $\mathcal{P}_{+}(t)$ then decreases with time except for two time intervals where 
this probability increases. 
The first time interval still survives with dissipation since the decoherence process leads to behave bosons as distinguishable particles 
(as seen in Figure \ref{fig: bf_eta} for $\eta = 0$). For fermions, the anti-symmetric wave function makes 
$ \mathcal{P}_{-, \text{pp}}(t) $ be 
higher than for bosons and backflow is then completely suppressed, both in the non-dissipative and dissipative regimes 
in spite of  interference. In other words, 
one can mathematically understand these behaviors by looking at the time slope of $\mathcal{P}_{\pm,\text{pp}}(t)$ at t=0. 
For our chosen parameters, $ \dfrac{d}{dt}\mathcal{P}_{+,\text{pp}}(t) \bigg|_{t=0} < 0 $ while 
$ \dfrac{d}{dt}\mathcal{P}_{-,\text{pp}}(t) \bigg|_{t=0} >0 $ confirming the appearance of the first backflow interval only for bosons.
It seems that the interference terms in the boson case favor this effect as a consequence of the corresponding symmetric wave function.
On the contrary, these interference terms together with the anti-symmetric character of the wave function do not lead to 
backflow in any regime.

\begin{figure}
	\centering
	\includegraphics[width=8cm,angle=-0]{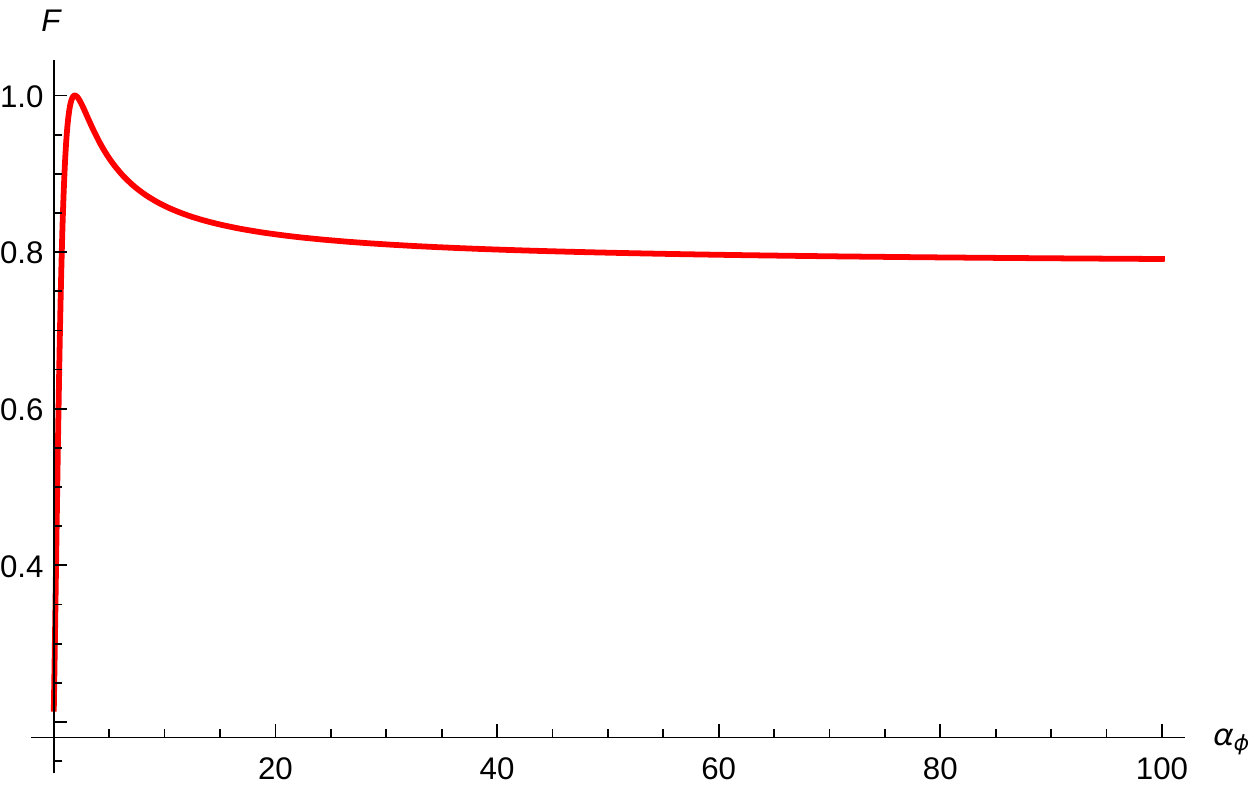}
	\caption{
		Fidelity $F$ of one-particle states $\chi$ and $\phi$ versus $\al_{\phi}$. Parameters are chosen as $ \si_p=0.05 $,  
		$ p_{0b}=0.3 $, $ p_{0a}=1.4 $, $\theta=\pi$, $\al_{\chi}=1.9$ (see text for symbols).
	} 
	\label{fig: fidelity} 
\end{figure}

\begin{figure}
	\centering
	\includegraphics[width=12cm,angle=-0]{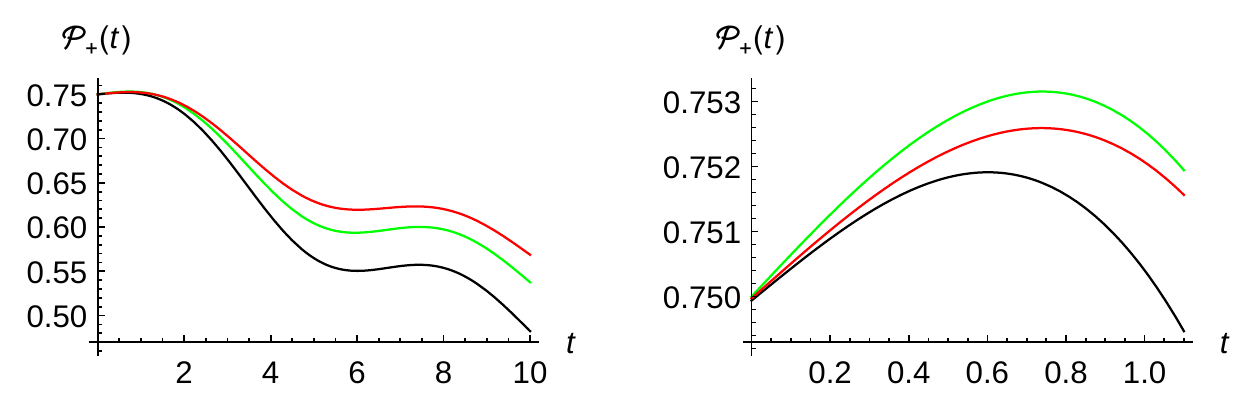}
	\caption{
		In the left panel, $ \mathcal{P}_+(t) $ versus time for the non-dissipative dynamics and for $ \al_{\phi}=1 $ (black curve), 
		$\al_{\phi}=1.9$ (green curve) and $\al_{\phi}=3.5$ (red curve). The right panel is a magnification of the first backflow.
		Parameters are chosen as $ \si_p=0.05 $,  $ p_{0b}=0.3 $, 
		$ p_{0a}=1.4 $, $\theta=\pi$ and $\al_{\chi}=1.9$. 
	} 
	\label{fig: bf_2p_B} 
\end{figure}

Finally, it is very illustrative to study backflow as a function of the fidelity of the initial one-particle wave functions $\chi$ and $\phi$. 
Fidelity of two pure states is defined as the square of the overlap between the states: $ F = | \la \chi | \phi \ra |^2$. 
The overlap between these states is given by
\begin{eqnarray} \label{eq: overlap}
\la \chi | \phi \ra &=& N_{\chi} N_{\phi} \left[ 1 + \al_{\chi} \al_{\phi} e^{-(p_{0a}-p_{0b})^2/8\si_p^2} ( e^{-i\theta} \al_{\chi} + e^{i\theta} \al_{\phi}) \right] 
\end{eqnarray}
where we have assumed that both $\chi$ and $\phi$ states have the superposed form given by Eq. (\ref{eq: wf_sup_t}) but this time 
with the same value of $\theta$ but  different values of $\alpha$; $\al_{\chi}$ and $\al_{\phi}$, respectively.
In Fig. \ref{fig: fidelity}, fidelity is plotted versus $\al_{\phi}$ for a given value of $\al_{\chi}$. As one expects when 
$\al_{\phi} = \al_{\chi}$, fidelity takes its maximum 
value i.e., becomes unity. As this figure shows, $ F $ becomes constant taking the value $ \approx 0.79$ for large values of $\al_{\phi}$.
Figure \ref{fig: bf_2p_B} displays $ \mathcal{P}_{+}(t) $ (for bosons) in the negative half-space for the free non-dissipative dynamics case.
In particular, in the left panel, $ \mathcal{P}_+(t) $ is plotted versus time for $ \al_{\phi}=1 $ (black curve), $\al_{\phi}=1.9$ (green curve) 
and $\al_{\phi}=3.5$ (red curve). The right panel is a magnification of the first backflow.  Parameters are chosen to be $ \si_p=0.05 $,  
$ p_{0b}=0.3 $, $ p_{0a}=1.4 $, $\theta=\pi$ and $\al_{\chi}=1.9$. 
As the right panel explicitly shows, the amount of backflow is maximum when the fidelity is also maximum, $\al_{\phi}=1.9$. 
Note that for the highest value of
fidelity both one-particle states are the same, and thus $ \Psi_+(x_1, x_2, t) = \chi(x_1, t) \chi(x_2 ,t) $ which this product has the form 
adopted for distinguishable particles. Additional calculations (not plotted here) show that for very small values of fidelity, 
backflow is not seen even for bosons. 

\section{Concluding remarks}

Backflow is a quite astonishing effect in quantum mechanics but it is curiously far less known than, for example, the tunneling effect. 
At least two reasons which could be argued are: first, it has not been experimentally observed yet and second, no clear application has 
been still devised.
However, at the fundamental level, the appearance or not of backflow can be a good check to see the degree of quantumness of a given system, 
either closed or open.
Although one can see non-zero amount of backflow in the limit $ \hb \rightarrow 0 $, Bracken and Melloy \cite{BrMe-JPA-1994} stated that 
this effect is a characteristic of the quantum mechanical description in terms of complex wave function with no classical analogue.  

As an extension of a previous work \cite{MoMi-EPJP-2020-2}, we have tackled a similar theoretical analysis 
dealing with  stretching Gaussian wave packets and  systems of two identical spinless particles  by adding a new ingredient
to this dynamics, the symmetry of the wave function. The role played by the 
interference terms and the corresponding symmetry of the two identical, spinless particles are crucial for this effect.
Bosons display backflow even for the dissipative case where the decoherence process leads to behave bosons  as 
distinguishable particles. On the contrary,  fermions do not exhibit backflow even in the non-dissipative regime in spite of displaying
interference terms; the anti-symmetric character of the wave function is strong enough to prevent it.
Strictly speaking, in our theoretical analysis, we have a set of parameters to choose freely. We can not affirm then that these results are going 
to be general, independent on any set of these parameters. However, we have shown at least that for a given set of parameters,
fermions do not exhibit backflow unlike bosons.  
For bosons, backflow has  also been analyzed in terms of fidelity which is a well-known property of two pure one-particle states. 
At very small values of fidelity, this effect is not seen even for bosons. In our opinion, extension of this type of analysis to 
more general dissipative frameworks like the Schr\"odinger-Langevin one, the use of other types of wave packets  as well as 
a systematic analysis of the initial parameter space are  necessary in order to acquire a more complete understanding of backflow
and try to confirm or not  the general validity of the results found here.

\vspace{1cm}
\noindent
{\bf Acknowledgements}

SVM acknowledges support from the University of Qom and SMA support from the Ministerio de Ciencia, Innovaci\'on y Universidades 
(Spain) under the Project FIS2017-83473-C2-1-P.



\end{document}